\documentclass[aps,pre,preprint,showpacs]{revtex4}

\usepackage{graphicx}% Include figure files

\begin{document}

%\preprint{APS/123-QED}

\title{The Effect of Sensory Blind Zones on Milling Behavior in a
Dynamic Self-Propelled Particle Model}

\author{Jonathan P. Newman}
\affiliation{Department of Bioengineering, Georgia Institute of Technology,
Atlanta, Georgia, 30332, USA}
\email{jnewman6@gatech.edu}
\author{Hiroki Sayama}
\affiliation{Department of Bioengineering, Binghamton University,
State University of New York, Binghamton, New York 13902, USA}
\affiliation{New England Complex Systems Institute, Cambridge,
Massachusetts 02138, USA}
\email{sayama@binghamton.edu}

\begin{abstract}
Emergent pattern formation in self-propelled particle (SPP) systems is
extensively studied because it addresses a range of swarming phenomena
which occur without leadership. Here we present a dynamic SPP model in
which a sensory blind zone is introduced into each particle's zone of
interaction. Using numerical simulations we discovered that the
degradation of milling patterns with increasing blind zone ranges
undergoes two distinct transitions, including a new, spatially
non-homogeneous transition that involves cessation of particles'
motion caused by broken symmetries in their interaction fields. Our
results also show the necessity of nearly complete panoramic sensory
ability for milling behavior to emerge in dynamic SPP models,
suggesting a possible relationship between collective behavior and
sensory systems of biological organisms.
\end{abstract}


\maketitle

Self-organization and pattern formation in self-propelled particle
(SPP) systems has been a topic of great interest in theoretical
physics, mathematical biology and computational science \cite{1,2}. It
is well understood that the emergence of cohesive swarming motions
requires neither leaders nor globally enforced organizational
principles. Various SPP models have been used to explore stability and
phase transitions of swarming patterns in response to varying noise
levels \cite{6,7,3} and other control parameters \cite{5,8,9,11}, as
well as to characterize distinct regimes in the parameter space
\cite{8,10,11,12}. Effort has also been allotted for addressing
biological questions concerning swarming behavior
\cite{9,13,14,15,5,24,25} and for designing non-trivial swarming
patterns from combinations of different kinetic parameter sets
\cite{17}.

SPP models may be classified into two distinct categories: kinematic
and dynamic \cite{12}. Kinematic SPP models typically assume that each
particle maintains a speed and orientation in accord with its local
neighbors \cite{2}.  In these models, particles cannot halt because
they are supplied with a minimal or constant absolute velocity
\cite{2,3,13,14}.  Kinematic models have been used for computational
modeling of collective behavior of constantly moving groups, such as
bird flocks, often implementing empirically constructed complex,
spatially discrete interaction zones and behavioral rules that reflect
perceptional or locomotive properties of the species being modeled
\cite{14,13,15,24,25}. On the other hand, dynamic SPP models describe
the motion of particles using differential equations based on
Newtonian mechanics that involve self-propulsion and pairwise
attraction/repulsion forces \cite{7}. It is known that such
models may robustly form coherent milling patterns from initially
random conditions even without explicit alignment rules
\cite{6,7,10,11,12}. Dynamic models have been used for both analytical
and numerical research on collective behavior of interacting particles
in general, with minimal complexity assumed in particles' intrinsic
behaviors.

Here we consider a new dynamic SPP model in which a sensory blind zone
is introduced into each particle's zone of interaction. Although the
assumption of sensory blind zones has been widely adopted in
kinematic SPP models \cite{13,15,14,24,25}, it has not been considered
within a dynamic framework. We specifically examine the effect of the
sensory blind zones on coherent milling behavior in dynamic SPP
models. In doing so, we discovered a novel transition that occurred
with an increasing range of blind zones, and found the system to be
highly sensitive to this type of perturbation.

\begin{figure}
\includegraphics[width=0.7\columnwidth]{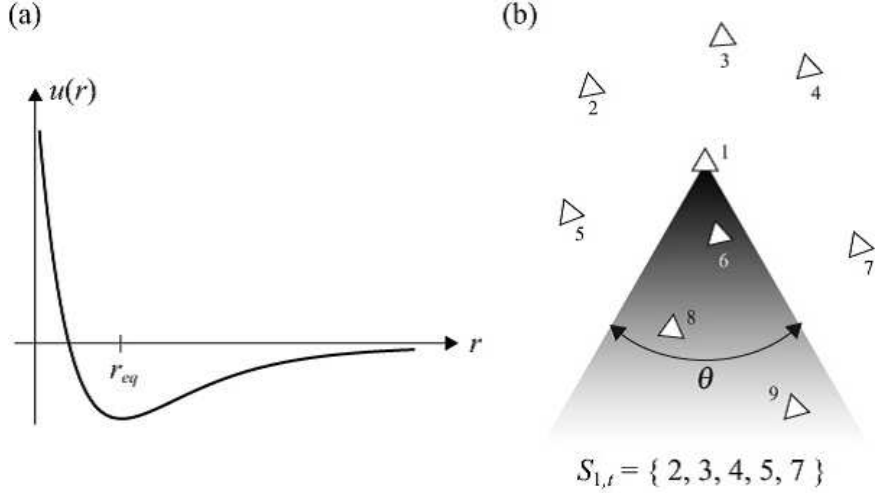}
\caption{Model assumptions used in the SPP model employed by this
study. (a) Shape of the pairwise interaction potential function $u(r)$
defined by Eq.~(\ref{pairwisepotential}), where $r$ is the distance
between two particles. This study explores the biologically relevant
regime of parameter settings, in which particles will accelerate away
from neighbors who are closer than, and toward neighbors further than,
the equilibrium distance $r_{eq} \equiv \frac{l_a
l_r}{l_a-l_r}\log\frac{C_r l_a}{C_a l_r}$ ($\approx 1.39$ with
parameter settings used in this paper). (b) A sensory blind zone
oriented opposite the direction of forward motion of the particle with
angular range $\theta$. Particles are represented by small
triangles. In this example, particles 6, 8 and 9 are within particle
1's blind zone, so their indices are not included in the set $S_{1,t}$
when generating repulsive/attractive forces acting on particle 1.}
\label{fig1}
\end{figure}

Our model describes the movement, within an open, two-dimensional,
continuous space, of $N$ self-propelled particles driven by soft-core
interactions whose dynamics are given by
\begin{eqnarray}
\frac{d x_i}{d t} &=& v_i , \\
m \frac{d v_i}{d t} &=& (\alpha - \beta |v_i|^2) v_i - \nabla U_i(x_i) , \label{eq2} \\
U_i(x) &=& \sum_{j \neq i} u(|x - x_j|) , \\
u(r) &=& C_r e^{-r/l_r} - C_a e^{-r/l_a} \quad (r \geq 0), \label{pairwisepotential}
\end{eqnarray}
where $x_i$ and $v_i$ are the position and the velocity of the $i$-th
particle ($i=1 \ldots N$), respectively; $m$ the unit mass of one
particle; $\alpha$ and $\beta$ the coefficients of propulsion and
friction, respectively; $U_i(x)$ the interaction potential surface for
the $i$-th particle; $u(r)$ the pairwise interaction potential
function (Fig.~\ref{fig1} (a)); $C_r$ and $C_a$ the amplitudes of
repulsive and attractive pairwise interaction potentials,
respectively; and $l_r$ and $l_a$ the characteristic ranges of
repulsive and attractive pairwise interaction potentials,
respectively. Eq.~(\ref{eq2}) includes a velocity-dependent locomotory
term and an interaction term achieved through a generalized Morse
pairwise interaction potential. For $\alpha, \beta > 0$, particles
will rapidly approach equilibrium velocity of magnitude $v_{eq} \equiv
\sqrt{\alpha / \beta}$ and the system will converge toward a structure
for which total dissipation is zero and particles are driven only by
conserved forces \cite{11}. This rule set has been employed, with some
mathematical variation, by many previous studies \cite{10,11,12,6}. In
this study, the shape of the pairwise interaction potential falls
within the biologically relevant regime defined as $C_r / C_a > 1$ and
$l_r / l_a < 1$, as described by \cite{11,12}.  In the biologically
relevant regime, individuals tend to move toward other individuals
that are further from, and away from individuals that are closer than,
some critical distance from themselves. This rule is generally
applicable to the kinetics of many different biological species and
natural systems \cite{1,15,19}.

We compare two experimental parameters in the above model: the
magnitude of stochastic force (noise) $\gamma$ and the range of
sensory blind zones $\theta$, the former analyzed in \cite{6} and the
latter our original extension. Sensory blind zones are incorporated
into the design of each particle to mimic the abilities of anisotropic
sensory systems observed in nature, such as vision. A sensory blind
zone is assumed to exist for each particle with an angular range
$\theta$ in a direction opposite to the direction of forward motion
(Fig.~\ref{fig1} (b)).

The inclusion of these parameters introduces discrete events into the
model, i.e., abrupt changes of velocity by stochastic force and entry
and exit of other particles into/out of sensory blind
zones. Consequently, we revised the equations of motion using discrete
time steps. The difference equations used for numerical simulation are
\begin{eqnarray}
\frac{x_{i,t+\Delta t} - x_{i,t}}{\Delta t} &=& v_{i, t+\Delta t}, \\
m \frac{v_{i,t+\Delta t} - v_{i,t}}{\Delta t} &=& (\alpha - \beta |v_{i,t}|^2) v_{i,t}
- \nabla U_{i,t} (x_{i,t}) + \gamma \xi_{i,t}, \\
U_{i,t}(x) &=& \sum_{j \in S_{i,t}} u(|x-x_{j,t}|) , 
\end{eqnarray}
where $x_{i,t}$ and $v_{i,t}$ are the position and the velocity of the
$i$-th particle at time $t$, respectively; $\xi_{i,t}$ a randomly
oriented vector with length 1 whose orientation is independent for
each evaluation; $U_{i,t}(x)$ the interaction potential surface for
the $i$-th particle at time $t$; and $S_{i,t}$ the set of indices of
all the particles whose positions are outside the blind zone of the
$i$-th particle at time $t$ (Fig.~\ref{fig1} (b)).

We conducted numerical simulations of this model to produce a milling
pattern similar in structure to those witnessed in schools of teleost
fish, insects, and microorganisms
\cite{1,11,12,8,9,18,15}. Specific values of fixed parameters are
as follows: $m = 1.0$, $C_r = 1.0$, $C_a = 0.5$, $l_r =0.5$, $l_a =
2.0$, $\alpha = 1.6$, $\beta = 0.5$. Initial conditions of each
simulation were such that particles were randomly distributed within a
square area of side length $2 l_a$, and each particle was randomly
oriented with magnitude of velocity randomly chosen from $[0, v_{eq}]$
as described in \cite{11,12,10}.

The model equations were numerically simulated from $t = 0$ to 200 at
interval $\Delta t = 0.01$. No spatial boundaries were enforced. For
simulations recording the effect of stochastic force, $\gamma$ was
varied from 0 to 10 at interval 0.5, while $\theta = 0$. For
simulations testing the effect of sensory blind zones, $\theta$ was
varied from 0 to $0.2 \pi$ at interval $0.01\pi$, while $\gamma =
0$. Each parameter setting was simulated using several population
sizes, $N = 200$, 300, 400, and 500. Ten simulation runs were
conducted for each condition.

Several metrics were used to characterize the simulation
results. These include average absolute velocity $V_{abs}$, ratio of
halting particles $H$, normalized angular momentum $M$, and normalized
absolute angular momentum $M_{abs}$, defined as follows (same or
similar metrics were used in \cite{11,12,15}):
\begin{eqnarray}
V_{abs} &=& \sum_i |v_i| / N \\
H &=& \left|\left\{ i, {\rm ~s.t.~} |v_i| < \mu v_{eq} \right\}\right| / N \\
M &=& \frac{|\sum_i r_i \times v_i|}{\sum_i |r_i||v_i|} \\
M_{abs} &=& \frac{\sum_i |r_i \times v_i|}{\sum_i |r_i||v_i|}
\end{eqnarray}
Here $r_i \equiv x_i - x_c$ where $x_c$ is the swarm's center of
mass. To measure $H$ we used 20\% of the equilibrium velocity
($\mu=0.2$) as a threshold to determine whether a particle was halting
or not. When used comparatively, $M$ and $M_{abs}$ make it possible to
distinguish single mill formation from double mill formation in which
two mills rotate with opposite sense around similar, but not
identical, centers of mass \cite{11,12}. All the metrics were averaged
over the last 10 time steps of each simulation.

\begin{figure}
\includegraphics[height=0.75\textheight]{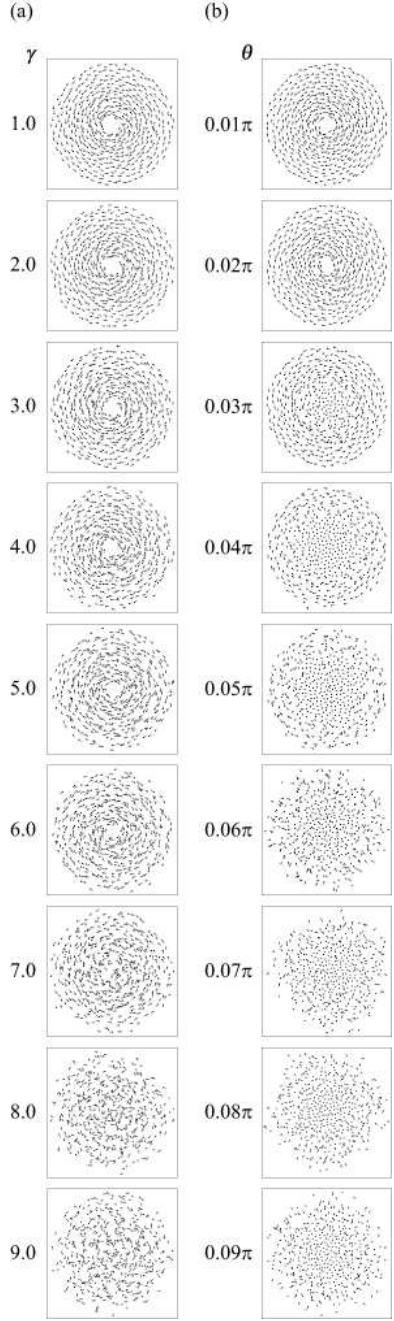}
\caption{A visual comparison of the effects of increasing stochastic
force and the effects of increasing range of sensory blind zones on
the milling behavior of 500 particles. Each image is a final snapshot
of a simulated particle swarm taken at $t=200$. Particles have tails
that represent the orientation and magnitude of their velocity. (a)
Results with increasing stochastic force $\gamma$ while
$\theta=0.0$. Transition from milling to disordered states occurred at
$\gamma \approx 7.0$. (b) Results with increasing range of sensory
blind zones $\theta$ while $\gamma=0.0$. Transitions from milling to
carousel and from carousel to surface disordered states occurred at
$\theta \approx 0.03\pi$ and $\theta \approx 0.06\pi$,
respectively. See also Fig.~\ref{fig3}.}
\label{fig2}
\end{figure}

\begin{figure}
\includegraphics[width=0.8\columnwidth]{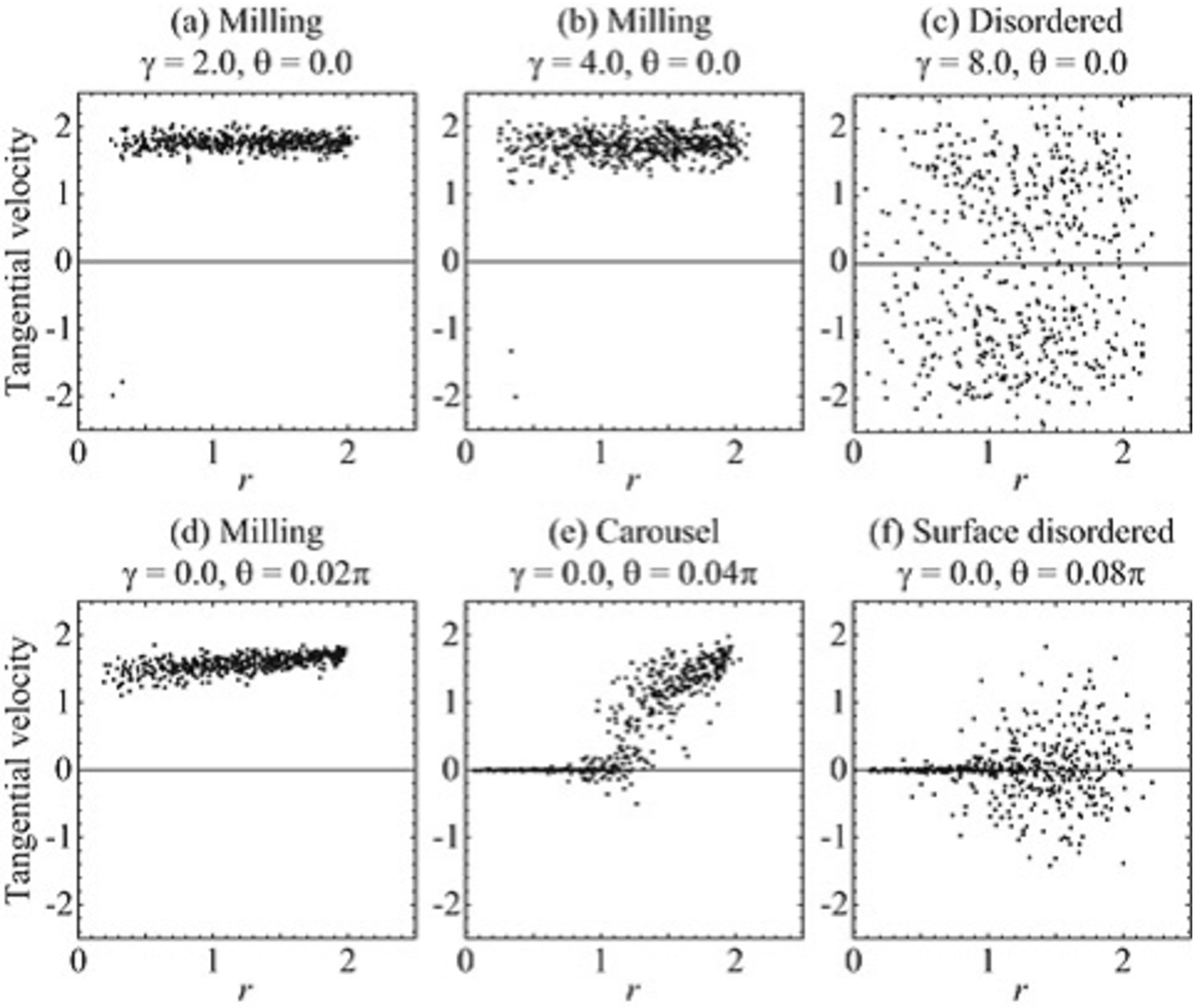}
\caption{Tangential velocities of particles at a distance $r$ from the
center of mass. Data were obtained from numerical simulations of 500
particles at $t=200$. The direction of rotation of the majority was
taken as positive.}
\label{fig3}
\end{figure}

Figs.~\ref{fig2} and \ref{fig3} depict the processes of structural
decay produced by stochastic force and blind zone perturbations on the
milling behavior of 500 particles. A transition from milling state to
disordered state was induced by increasing the magnitude of stochastic
force across $\gamma \approx 7.0$ (Fig.~\ref{fig2} (a),
Fig.~\ref{fig3} (a)--(c)).  When the range of sensory blind zones
$\theta$ was increased, however, structural degradation was very
different, involving a new spatially heterogeneous transition
(Fig.~\ref{fig2} (b), Fig.~\ref{fig3} (d)--(f)). Initiation of
collapse occurred at $\theta \approx 0.03 \pi$, where particles near
the center of the mill ceased rotation and formed a stationary core
that has not previously been described. We call this new state a
``carousel'' state. This state is different from the rigid-body
rotation reported in \cite{11,12} and the compact but disordered state
of \cite{6} because it is characterized by a sharp boundary between
the milling surface and the central stationary core made of particles
with near zero velocity (Fig.~\ref{fig3} (e)). A secondary transition
was observed across $\theta \approx 0.06 \pi$ where the particles
moving in the periphery became abruptly disordered and lost coherence
in motion while the particles in the central core remained stationary
(Fig.~\ref{fig3} (f)). We call this concluding state a ``surface
disordered'' state.

Particles inside the core in the carousel and surface disordered
states lose their velocity due to a perceptual and consequent force
asymmetry. A sensory blind zone creates a longitudinal imbalance
between forces derived from particles ahead and forces from particles
behind. Because the pairwise equilibrium distance $r_{eq}$ is much
longer than the characteristic distance between neighboring particles
in a swarm, the imbalance takes effect in the regime of repulsive
interactions and thus results in a net resistance against
self-propulsion of particles. A particle near the center of the swarm
has more particles to its front and back than does a particle rotating
in the periphery, since the density of particles is inversely related
to the distance from the center of the swarm under the parameter set
used here \cite{10} (numerically confirmed in our simulation results;
data not shown). Thus, the net resistance against forward motion
resulting from the blind zone is larger for particles rotating close
to the mill's center. Within a certain distance to the mill's center,
the resistance exceeds the range of self-propulsive force possible in
Eq.~(\ref{eq2}), and consequently particles cease motion. This
transition does not occur due to increasing $\gamma$ because the
effect of stochastic force is spatially isotropic: it is equally
likely to force a particle in any direction and therefore does not
lead to cessation of movement.

\begin{figure*}
\includegraphics[width=0.75\columnwidth]{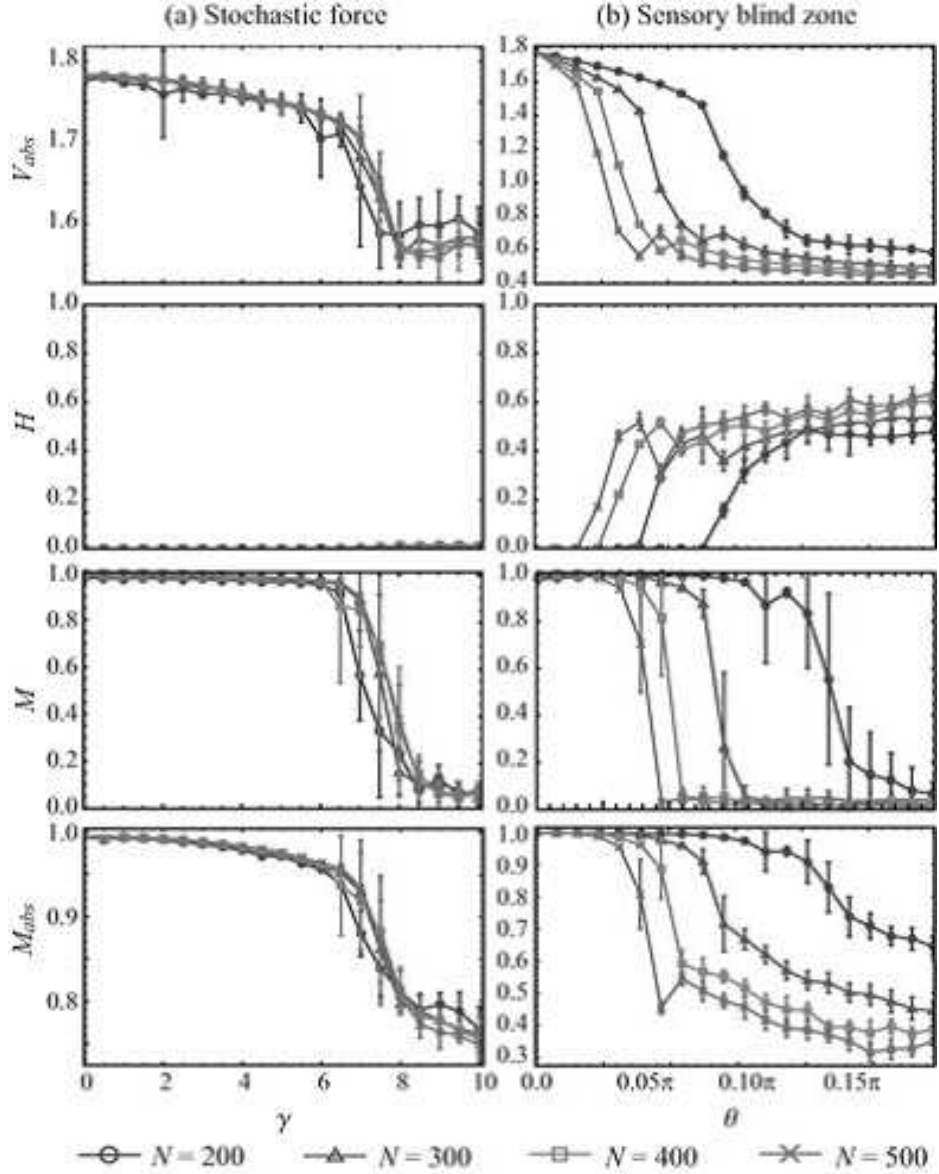}
\caption{(Color online) Comparison of the values of the four metrics
(average absolute velocity $V_{abs}$, ratio of halting particles $H$,
normalized angular momentum $M$ and normalized absolute angular
momentum $M_{abs}$) measured for over all simulations for $N$ = 200,
300, 400, 500. Each data point represents the average of ten
simulation runs with an error bar, measured in standard
deviations. (a) Results with increasing stochastic force, where the
collapse of milling behavior is always reached at $\gamma \approx 7.0$
regardless of $N$. (b) Results with increasing range of sensory blind
zones. The collapse of mill behavior is twofold; the top two plots
($V_{abs}$ and $H$) capture the first transition from milling to
carousel, while the bottom two ($M$ and $M_{abs}$) capture the second
transition from carousel to surface disordered. The critical values of
$\theta$ for these transitions depend on $N$, indicating that larger
populations are more susceptible to blind zone perturbations.}
\label{fig4}
\end{figure*}

Fig.~\ref{fig4} summarizes all the simulation results, showing the
dependence of the final values of $V_{abs}$, $H$, $M$ and $M_{abs}$ on
$\gamma$, $\theta$ and $N$. The onset and the mechanism of structural
degradation of milling behavior are different between cases with
increasing $\gamma$ and $\theta$. The milling structure is fairly
robust to small $\gamma$, and it suddenly collapses at $\gamma \approx
7.0$, nearly independently of $N$. The $H$ plot shows no particles
halting in this transition. In contrast, plots of increasing $\theta$
illustrate that structural degradation in increasing $\theta$ is a
two-fold process. The first transition from milling to carousel was
detected in $V_{abs}$ and $H$ (emergence of halting particles and
consequent decrease of average velocity). The second transition from
carousel to surface disordered was detected in $M$ and $M_{abs}$ (loss
of coherence in angular momentum). It was also observed in our results
that the onsets of these transitions depended significantly on
$N$. This can be understood in that larger $N$ increased the density
at the mill's center and made particles more reactive to blind zone
induced halting.

The blind zone ranges used in these simulations were extremely small
from a biological viewpoint. The largest value tested, $\theta =
0.2\pi$, was just 10\% of the perception range, which was more than
sufficient to destroy milling patterns in all cases.  Kinematic SPP
models, on the other hand, can produce and maintain milling patterns
despite considerable sensory blind zones \cite{15}. To understand this
discrepancy in model properties, we note one important difference
between kinematic and dynamic frameworks: while particles in kinematic
models are always constrained to move with a non-zero velocity, there is a possibility
for particles to halt in dynamic models that becomes significant in
the presence of rear blind zones like those assumed in our model. This
leads us to a hypothesis that milling behavior in an aggregate of
organisms may sensitively depend on their ability to maintain constant
velocity. Specifically, for organisms that keep moving autonomously at
a near constant pace, milling behavior emerges relatively easily even
with considerable sensory blind zones. In contrast, for organisms
whose motion strongly depends on (either sensory or physical)
environmental stimuli, milling behavior requires a nearly complete
panoramic range of interaction, especially to perceive the pressure
from behind and gain enough forward propulsion to maintain constant
velocity.

There are several biological observations that directly or indirectly
support our hypothesis. Milling behavior is often reported in groups
of microorganisms and insects \cite{10,11,9,1,3,5,8,18}.  These
organisms rely chiefly on direct physical contact and chemical sensory
input, respectively, when forming mills and therefore use
omnidirectional sensory capabilities to form aggregates. They also
have the ability to cease motion. Additionally, there is a great deal
of evidence supporting the isotropic sensory ability of the lateral
line in teleost fish, some of which demonstrate milling behavior
\cite{21}.  A recent study that investigated the superficial organization of neuromasts composing the lateral line in goldfish showed
that neuromasts' most sensitive axes were oriented in almost every direction  \cite{22}. Moreover, a recent
study on Mormon crickets \cite{16} reports that the physical,
cannibalistic threat of protein and salt deprived individuals from behind plays
a critical role in creating a large-scale coherent march. When some
crickets are immobilized and therefore unable to respond to a push
from behind, the march halts.  This study provides clear evidence
supporting our conjecture that inputs (physical pressure) from behind
a particle can be important in the formation of coherent swarming
patterns.

In summary, we computationally studied the effects of sensory blind
zones on the stability of self-organizing mill formation in a dynamic
SPP model. We found that milling behavior collapses through two
spatially distinct transitions in response to an increasing range of
rear blind zones, characterized by a halting regime emanating from the
center of the swarm and then a disorganization of coherent motion in
the periphery area. This is quite different from pattern collapse
observed with increasing stochastic force described by a spatially
uniform transition to a compact but disordered state
\cite{6}. Combined with other results obtained with kinematic SPP
models, our results suggest a possible relationship between collective
behavior and sensory systems of biological organisms: species that
engage in mill formation in nature may necessarily have an
omnidirectional sensory system if they do not maintain constant
velocity by themselves. This is a hypothesis testable and falsifiable
through experimental observation.

\begin{acknowledgments}
We thank Boris Chagnaud, Kurt Wiesenfeld and Stefan Boettcher for
their insightful comments on this manuscript.
\end{acknowledgments}

\end{document}